\pdfoutput=1 % this has to be within the first five lines
\documentclass[5p]{elsarticle}

\usepackage{graphicx}
\usepackage{hyperref}
\usepackage{amsmath,amsfonts}

\begin{document}

%\preprint{\tt YGHP-12-52}

\title{Knotted domain strings}

\author{Minoru {\sc Eto}}
\ead{meto(at)sci.kj.yamagata-u.ac.jp}
\author{Sven Bjarke {\sc Gudnason}}
\ead{gudnason(at)phys.huji.ac.il}
\address{Department of Physics, Yamagata University,
    Yamagata 990-8560, Japan}
\address{Racah Institute of Physics, The Hebrew
    University, Jerusalem, 91904, Israel} 

\begin{abstract}
We construct meta-stable knotted domain strings on the surface of a soliton of
the shape of a torus in 3+1 dimensions. We consider the simplest case
of $\mathbb{Z}_2$ Wess-Zumino-type domain walls for which we can cover
the torus with a domain string accompanied with an anti-domain
string. 
In this theory, all $(p,q)$-torus knots can be realized as a linked
pair of a(n) (un)knotted domain string and an anti-domain string.
\end{abstract}


\maketitle

%%%%%%%%%%%%%%%%%%%%%%%%%%%%%%%%%%%%%%%%%%%%%%%%%%%%%%

\section{Knotted vortex strings}
More than 140 years ago,
Lord Kelvin proposed an interesting idea that atoms could be conceived  
as stable knotted vortex loops. Although this idea was not
successful as a theory of atoms, it led to the celebrated mathematical
knot theory today. 
Knots are one of the most fascinating structures frequently appearing
in Nature and they are found to be important in diverse areas of
physics such as high energy physics, cosmology and condensed matter
physics. 

It was a long-standing question whether a stable knotted structure
actually exists in a dynamical system.
Indeed, until quite recently, no stable knot structures were found.
In 1996, Gladikowski and Hellmund \cite{Gladikowski:1996mb} as well as 
Faddeev and Niemi \cite{Faddeev:1996zj} found stable knot-like
structures made of (stable) topological solitons, Hopfions, which are
(un)knotted closed loops of vortex tubes  in the Faddeev-Skyrme model
\cite{Faddeev:1975}.
The Faddeev-Skyrme model is an $O(3)$ nonlinear sigma model with the
addition of a (four-derivative) Skyrme term. 
With the aid of the recent drastically improved computer power, they
succeeded in constructing numerical solutions of Hopfions with small
charges. 
Faddeev and Niemi \cite{Faddeev:1996zj} conjectured that all torus
knots can be constructed from stable knotted vortex tubes. 
Soon after, Battye and Sutcliffe found beautiful higher-charged
Hopfions numerically, which have both link and knot structures
\cite{Battye:1998pe}.

After the discovery in terms of numerical solutions, the knotted
solitons have been studied extensively in the literature.
However, all knotted solitons, known so far, are obtained from 
closed vortex flux tubes. 
The purpose of this Letter is to demonstrate the existence of a
different type of knotted soliton, which lives on the surface of
toroidal structures in 3+1 dimensions. 
Viz.~we construct non-planar domain strings on the surface of a
ring-shaped soliton, like a vorton \cite{Davis:1988ij}, a
superconducting string loop (spring) \cite{Copeland:1987th} or a
$Q$-ring \cite{Axenides:2001jj} etc.

\section{The model}
In \cite{Bazeia:1999su,Brito:2001ga} it
was proposed that non-planar domain wall networks might exist on the
surface of a host soliton star which is a spherically compactified
domain wall. This idea was investigated numerically in
\cite{Sutcliffe:2003et} in a simple concrete model having two complex
scalar fields and $U(1)\times\mathbb{Z}_n$ global symmetry.
In \cite{Sutcliffe:2003et}, the host soliton (star) is a large
$Q$-ball \cite{Coleman:1985ki} and an attempt was made to tile the
surface of the $Q$-ball with almost-BPS planar domain-string networks
in the Wess-Zumino model \cite{Saffin:1999au}.
However, in \cite{Sutcliffe:2003et} it was proven that only a few
spherical polyhedra can be constructed on the surface of a sphere.

Although an attempt of tiling the domain-string network on a sphere
did not turn out very successful, we can still ask whether it is
possible to tile other Riemann surfaces with domain strings. 
Namely, how does the answer depend on the topology of the host
soliton. Because we are interested in a solitonic object which may be
dynamically produced, we choose a torus, $T^2$, which is one of the
simplest geometries. 
Indeed, many ring-shaped solitons are known, e.g.~a closed loop of the
superconducting cosmic strings \cite{Davis:1988ij}, springs
\cite{Copeland:1987th} and Hopfions \cite{Gladikowski:1996mb}. 

\begin{figure}[t]
\begin{center}
\includegraphics[width=8cm]{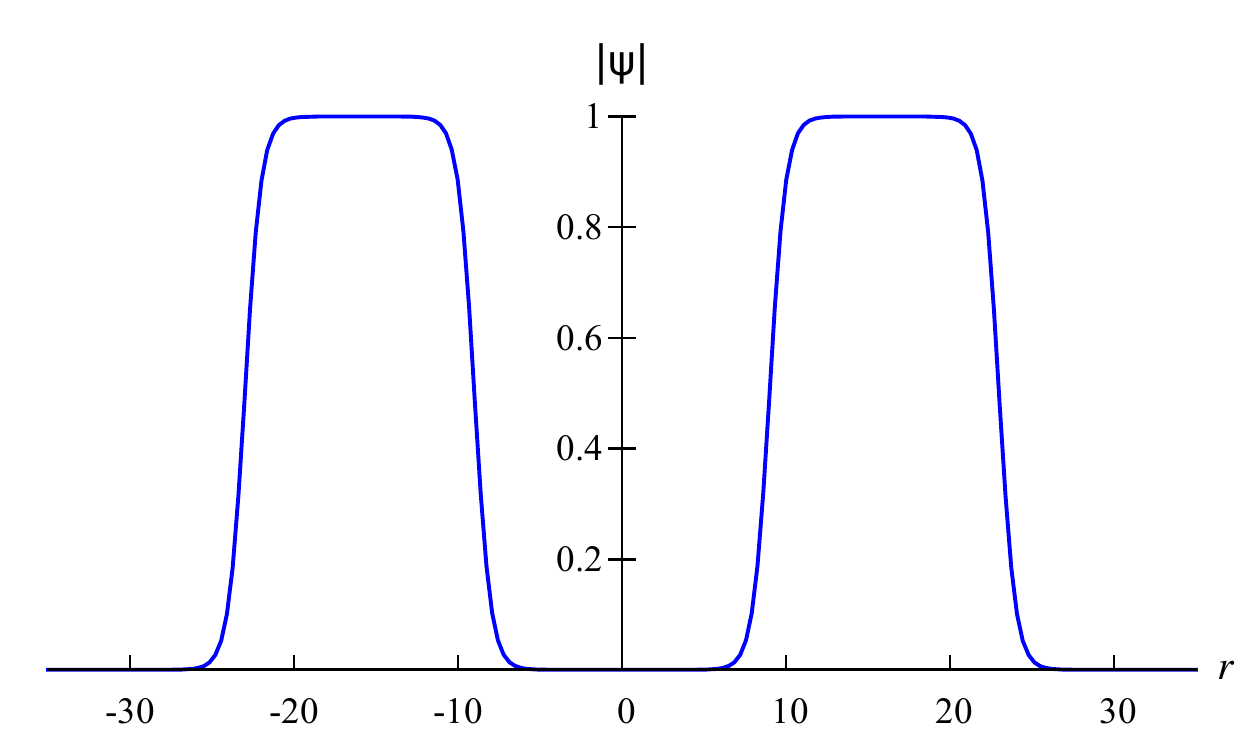}
\caption{The cross-section profile of $\psi$ for a ring-shaped soliton
  at $z=0$. The ring is axially symmetric with respect to the
  $z$-axis.} 
\label{fig:torus}
\end{center}
\end{figure}

\begin{figure}[ht]
\begin{center}
\includegraphics[width=8cm]{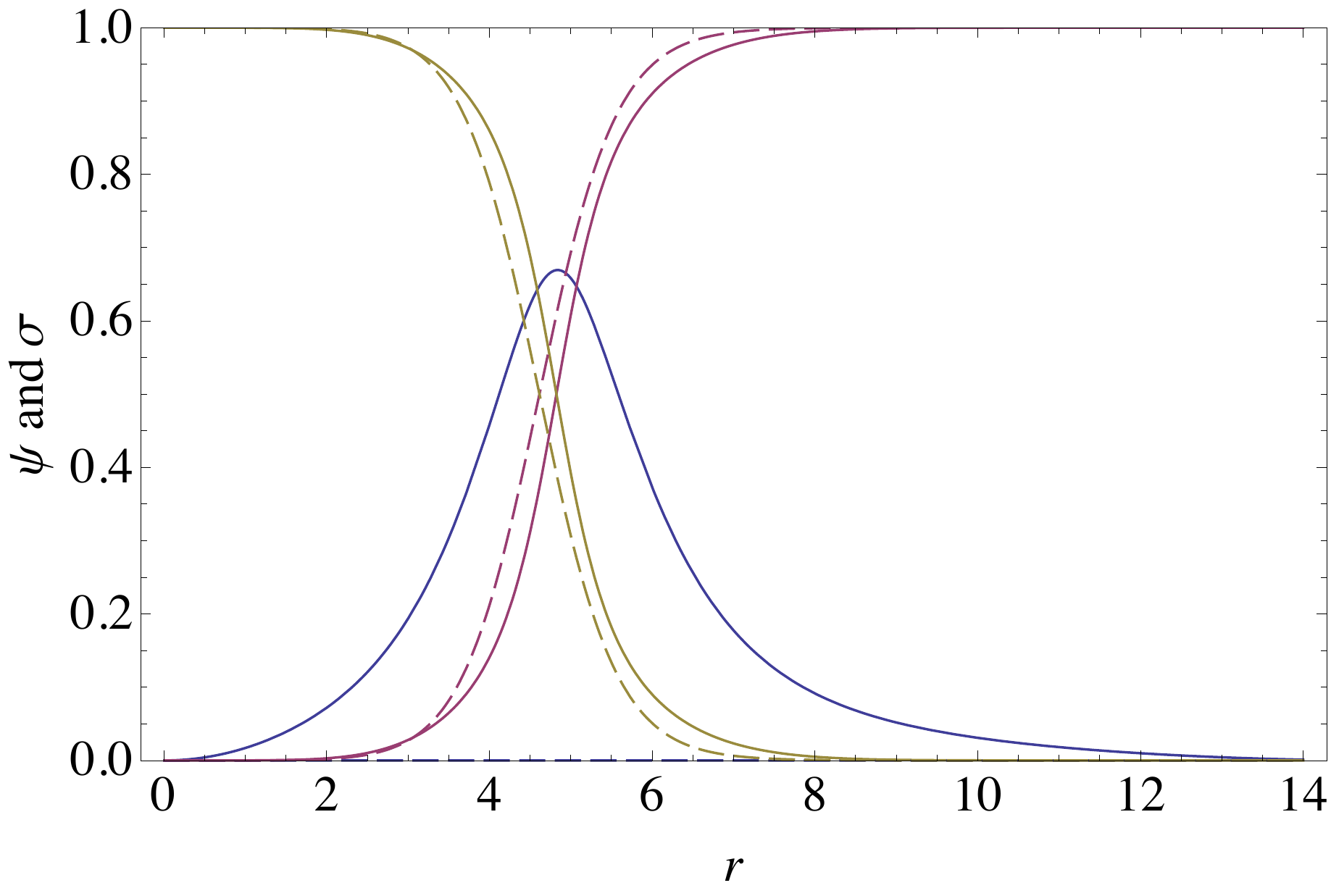}
\caption{The twisted vortex profile functions $|\psi(r)|^2$ (in
  gold/decreasing with $r$),
  $|\sigma(r)|^2$ (in violet/increasing with $r$) and $|\phi/v|^2$ (in
  blue) for $k=2$ with $\beta = 0$ (dashed curves) and $\beta=v=1$
  (solid curves). The parameters are chosen as $\gamma=1$, $\omega=1$
  and $p=1.1$. One can increase the toroidal radius
  $R_p$ by increasing the winding number $k$ in $\mathcal{L}_1$.} 
\label{fig:vorton}
\end{center}
\end{figure}

We will work in the framework of the simple model proposed in
\cite{Sutcliffe:2003et} which was used to construct the domain-string
networks on $Q$-balls. 
That is, we will use a single complex scalar field $\phi$ with an effective
potential of the Wess-Zumino type coupled to some other field which
belongs to a host configuration.
Hence, the Lagrangian contains two parts
\begin{eqnarray}
{\cal L} = {\cal L}_1[A_\mu,\psi_i] + {\cal L}_2[\psi_i,\phi],
\label{eq:model}
\end{eqnarray}
where $\mathcal{L}_1$ may include multiple scalar fields $\psi_i$
and gauge fields $A_\mu$ which are needed to form a stable ring
soliton. 
The Wess-Zumino field $\phi$ couples to $\psi_i$ only via
$\mathcal{L}_2$. 
We need not specify a particular model for $\mathcal{L}_1$, but we
require it to have a solitonic field configuration admitting a
ring-shaped profile function. 
The only condition which we require for one of the $\psi_i$'s, say
$\psi_1$, is that it is non-zero inside and zero outside of the
ring, see Fig.~\ref{fig:torus}. 
Indeed, this is a universal property of the condensate field for the
well-known vortons and springs. 
Then, for ${\cal L}_2[\psi_i,\phi]$, we consider the specific model
\begin{eqnarray}
{\cal L}_2 &=& \frac{1}{2}|\partial_\mu \phi|^2 
- \beta^2\left|\eta[\psi_1] - \frac{\phi^2}{v^2}\right|^2, \\
\eta[\psi_1] &\equiv& 4|\psi_1|^2\left(1-|\psi_1|^2\right).
\end{eqnarray}
As done in \cite{Sutcliffe:2003et}, we treat $\beta$ as a small
parameter, such that the profile of the torus configuration of 
${\cal L}_1$ receives only a negligible correction.
This can be seen as follows.
To be concrete, let us consider one of the simplest models as 
${\cal L}_1$ consisting of the complex fields 
$(\psi_1,\psi_2) = (\psi,\sigma)$:
\begin{eqnarray}
{\cal L}_1[\psi,\sigma] &=& |\partial_\mu \psi|^2 + |\partial_\mu \sigma|^2 
- \gamma |\psi|^2|\sigma|^2 \ , \\
\label{eq:L1_example}
\eta[\psi] &=& 4|\psi|^2|\sigma|^2 = 4|\psi|^2(1-|\psi|^2) \ ,
\end{eqnarray}
with the constraint $|\sigma|^2 + |\psi|^2 = 1$ and $\gamma$ takes on
a positive value. 
The stable vorton solution in this model ($\beta=0$) was obtained in 
\cite{Radu:2008pp}. 
The field $\sigma = f(r) e^{ik\theta}$ is called the vorton field, and
the condensate $\psi$ whose phase increases along the vorton as $\psi
= g(r) e^{i (\omega t + p z)}$, with $r,\theta$ being polar
coordinates. 
Since we just want to estimate how much the host soliton is modified
by turning on a non-zero coupling, $\beta$, we simplify the problem
here and consider a so-called twisted-vortex string \cite{Radu:2008pp}
on a periodic interval, $z = 2\pi/p$, instead of the vorton. 
Let us consider the twisted-vortex string in ${\cal L} = {\cal L}_1 +
{\cal L}_2$. 
Firstly, the vacuum structure of ${\cal L}$ is not changed with
respect to that of ${\cal L}_1$ in Eq.~(\ref{eq:L1_example}).
Furthermore, all the conserved charges are independent of $\beta$. 
The existence of the same kinds of (host) solitonic configurations in 
$\mathcal{L}$ with $\beta\neq 0$ as in ${\cal L}$ with $\beta = 0$
follows straightforwardly. 
Indeed, we obtained numerically a twisted vortex string solution which
is shown Fig.~\ref{fig:vorton}. 
In Fig.~\ref{fig:vorton}, we have chosen a relatively large value for
the coupling, i.e.~$\beta = 1$, but the deformation of the profile
functions, $\psi$ and $\sigma$, remains small.
Thus we conclude that the shape of the host ring is quite insensitive
to the $\beta$ coupling -- even for order one values -- and thus we
can safely ignore the back reaction as long as $\beta$ is kept
sufficiently small. 
There is, however, an important difference between $\beta = 0$ and $\beta
\neq 0$: $\phi \neq 0$ for the latter. 
Since $\eta[\psi_i]=0$ both inside and outside of the ring, $\phi$
only develops a non-zero VEV $\phi \simeq v$ near the surface of the
ring, in order to minimize the energy contribution from the potential
term.  
Clearly, there exists another configuration, namely $\phi \simeq - v$
near the surface of torus. 
These two configurations have exactly the same energy due to the
$\mathbb{Z}_2$ symmetry of the model. 
Hence, the $\mathbb{Z}_2$ symmetry is spontaneously broken by the ring
soliton, which gives rise to domain strings on the host ring soliton. 

Once we ignore the back reaction to the host soliton, ${\cal L}_2$ can
be seen effectively as the $2+1$ dimensional Wess-Zumino model with a
$\mathbb{Z}_2$ discrete symmetry, where the field $\phi$ has VEVs
$\langle\phi\rangle\simeq \pm v$.
As well known, the Wess-Zumino model in $2+1$ dimensions admits domain 
strings interpolating those two vacua having tension and transverse
size, respectively,
\begin{equation}
T = \frac{2^{5/2}}{3} \beta v, \quad
d = \frac{v}{\beta}.
\label{eq:tension}
\end{equation}
Therefore, we can take $\beta$ parametrically small keeping the size
of the domain string $d$ fixed. 
We will choose $\beta \simeq v$ such that $d$ is of order one in the
following, and hence the tension of the domain string is of order
$\beta^2$. 
Since, no physical parameters depend on only $\beta$, we can keep the
size of the domain strings large and choose a small value of
$\beta$. In this way the back reaction is parametrically negligible
and need not be a concern.

In the following, we will tile the surface of the ring soliton with
these two domains.

\section{Numerical calculation} 
We solved the partial differential equations (gradient flow equations)
with a finite difference  
method, more precisely using a Crank-Nicolson algorithm on a
$160^2\times 80$
square lattice times a relaxation time axis with Courant number,
$\Delta t/\Delta x_i = 0.2$:
\begin{equation}
- \partial_\phi \mathcal{E} 
= {\rm EOM}[\nabla\phi,\phi] = \partial_t \phi \, ,
\end{equation}
such that when the configuration does not change anymore, the soliton
configuration is obtained. This is the relaxation method.
The spatial lattice has the lengths $60^2\times 30$ and stepsize
$\Delta x_i = 0.375$ and we chose $\beta=v/3=0.1$ yielding
$d=\sqrt{3}$. Hence, the stepsize is small enough to resolve the
domain string (there is about $4.6$ lattice points on the domain
string itself). 
Since the Crank-Nicolson algorithm is implicit
\begin{equation}
\frac{1}{2}\left[
-\partial_\phi\mathcal{E}(t)
-\partial_\phi\mathcal{E}(t+\Delta t)\right] 
= \frac{\phi(t+\Delta t) - \phi(t)}{\Delta t} \, , 
\end{equation}
$\phi(t+\Delta t)$ needs to be calculated by means of solving a matrix 
equation for which we use a biconjugate gradients method. 
Since the equation of motion is non-linear, we linearize the equation
(only for $\phi$ at time $t+\Delta t$)
\begin{equation}
\phi = \phi_0 + \delta\phi \, ,
\end{equation}
i.e.~we keep the full non-linear expression for $\phi_0$ but truncate
$\delta\phi$ to linear order. We then iterate several times until the
solution of the next time slice, $\phi(t+\Delta t)$, converges. 
Thereafter, the routine is continued until the variation of the field
in time is small enough.

The Figs.~\ref{fig:10}--\ref{fig:21} are obtained by the above explained 
Crank-Nicolson algorithm.
As a check we have also carried out the same calculation using
Mathematica.
We obtain equally good solutions with the calculation
done in Mathematica, see Figs.~\ref{fig:32}--\ref{fig:43}. 

With an appropriate initial configuration, one obtains the desired
results as final states of the relaxation. 
If the configuration is unstable, it collapses to a single
vacuum. Since we obtained non-trivial domain strings by means of the
relaxation, they represent stationary points of the energy.

\section{Knotted domain strings}
As we will explain,
domain strings on the ring surface are nothing but torus
knots. Therefore, they are naturally characterized by a pair of 
co-prime integers $(p,q)$.
The number $p$ denotes the poloidal winding number (around the
meridian circle), while $q$ is the toroidal winding number (around the
longitudinal circle) of the torus.
As well known, torus knots are prime and chiral. Torus knots with
$pq>0$ are right-handed and $pq<0$ are left-handed. A $(p,q)$-knot is
identical to the $(-p,-q)$-knot.
\begin{figure}[ht]
\begin{center}
\includegraphics[width=8cm]{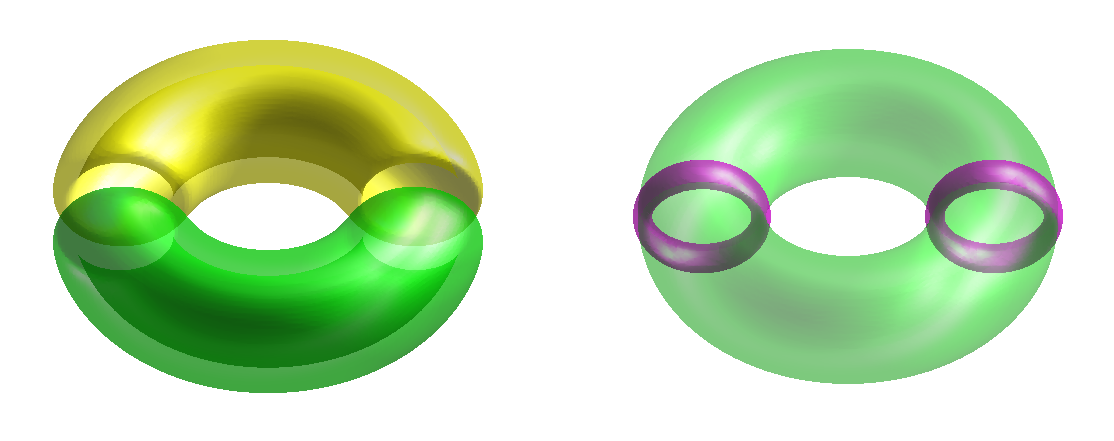}\\
\includegraphics[width=8cm]{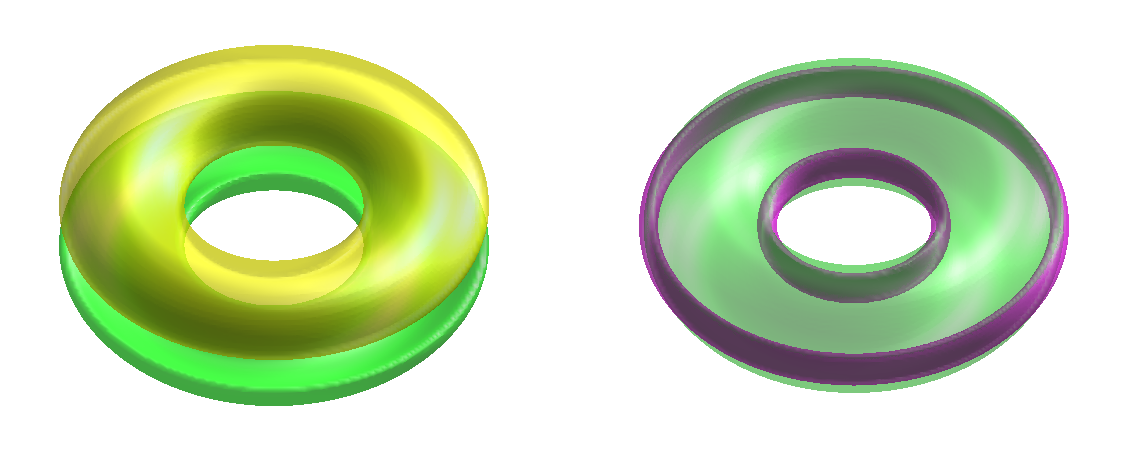}
\caption{The left panel shows the two domains. The two domain strings  
(i.e.~energy density isosurface) are shown in the right panel. The
  upper figures are for the $(1,0)$ type, i.e.~poloidal strings, while
  the lower figures are for the $(0,1)$ type, i.e.~toroidal strings.} 
\label{fig:10}
\end{center}
\end{figure}

The simplest configurations are the $(1,0)$ and $(0,1)$.
These have unknotted closed domain strings sitting on the
antipodal points of the torus, see Fig.~\ref{fig:10}. 
%(Fig.~\ref{fig:nb10} in Appendix). 
Clearly, the configuration $(0,1)$ is unstable against small
perturbations because the smaller string loop is preferred
energetically. Thus, the larger string will shrink and annihilate the 
smaller one since the net $\mathbb{Z}_2$ charge of the configuration
is trivial.

\begin{figure}[ht]
\begin{center}
\includegraphics[width=8cm]{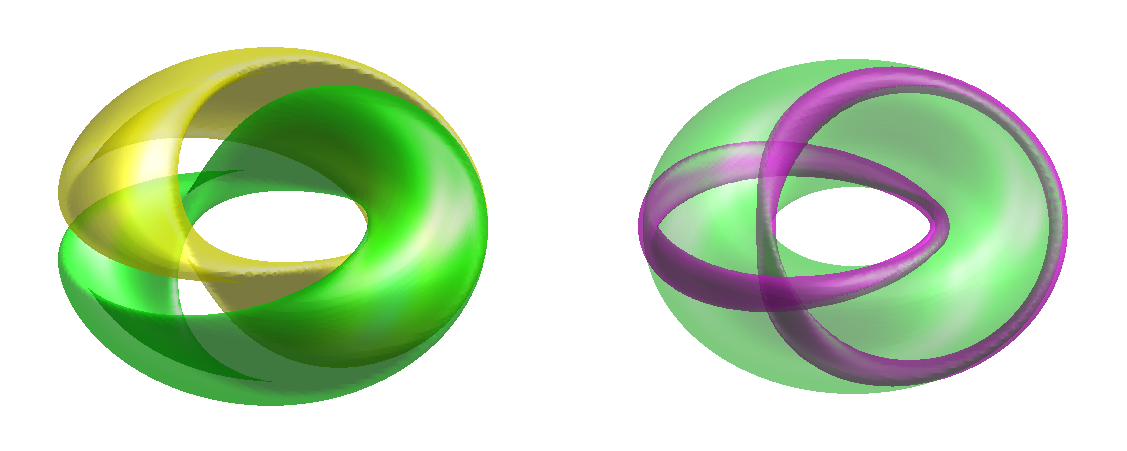}
\caption{Hopf link $(1,1)$: The left figure shows the tiling of the
  surface of a torus with two domains and the right figure shows that
  the domain strings are linked once.} 
\label{fig:11}
\end{center}
\end{figure}

The Hopf link appears for the $(1,1)$ type. The configuration is
unknotted but two domain strings are singly linked which is a
so-called Hopf link, see Fig.~\ref{fig:11}.
%(Fig.~\ref{fig:nb11} in Appendix).  
Each string loop winds both cycles of the torus once. The domain
string and anti-domain string sit near the antipodal point with
respect to the other on the torus.\footnote{For instance, for the Hopf
link (1,1), the string tension squeezes the strings a bit together
upon minimization of energy and thus at the point where the strings
are sitting orthogonal to the toroidal radius, they are not completely
antipodal to one another.} They are, however, nearly maximally
separated from each other in most of the configuration.

\begin{figure}[ht]
\begin{center}
\includegraphics[width=8cm]{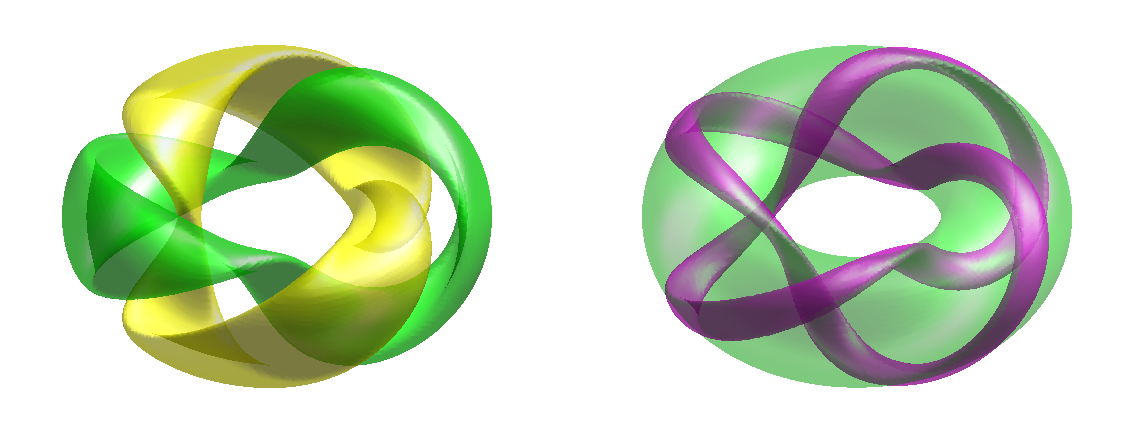}\\
\includegraphics[width=8cm]{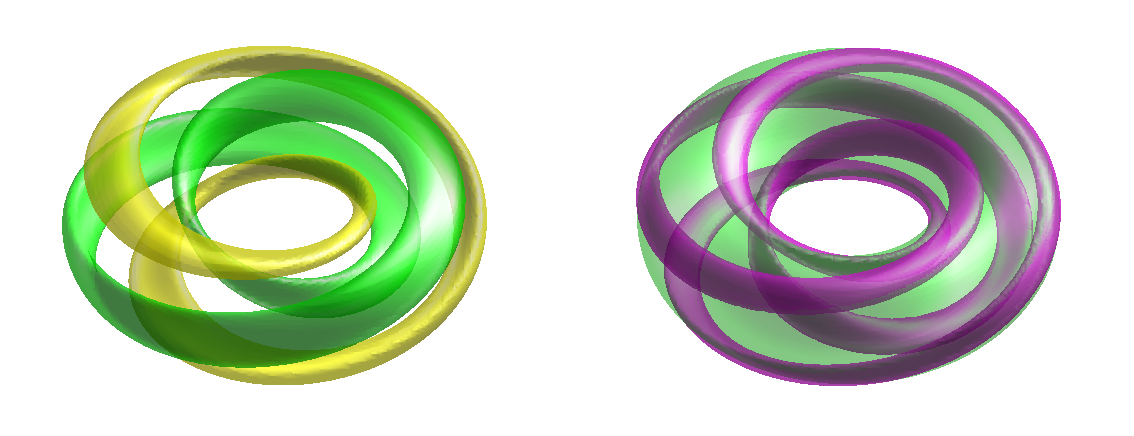}
\caption{Solomon's links: The upper figures show the $(2,1)$
  configuration and the lower ones show the $(1,2)$-type
  strings. These are unknotted but doubly linked configurations.} 
\label{fig:21}
\end{center}
\end{figure}

The unknotted but doubly linked strings are obtained for the $(2,1)$ and
$(1,2)$ cases. 
They are called Solomon's links, see Fig.~\ref{fig:21}.
%(Fig.~\ref{fig:nb21} in Appendix). 

\begin{figure}[ht]
\begin{center}
\includegraphics[width=7.5cm]{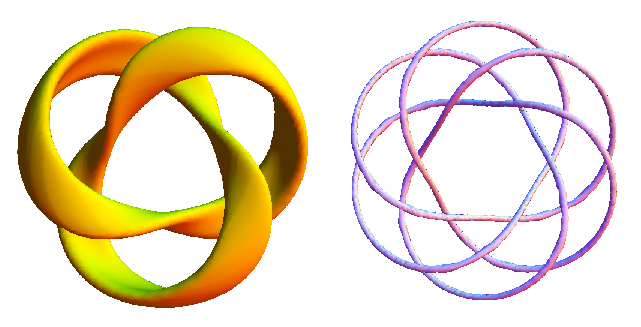}\\
\includegraphics[width=7.5cm]{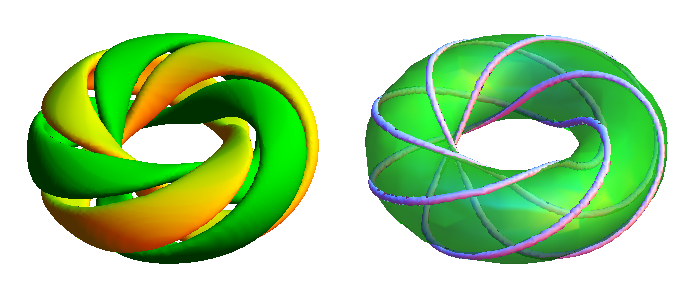}
\caption{The sixthly linked trefoil knot with winding numbers
  $(3,2)$. The top-left figure shows the region where 
  ${\rm Re}[\phi]>0$ which is a trefoil form. The top-right figure
  displays the two linked trefoil domain strings in the view from
  above. The bottom-left figure shows the two domains on the torus
  surface and the bottom-right figure is a 3D view of the trefoil
  knots.} 
\label{fig:32}
\end{center}
\end{figure}
\begin{figure}[ht]
\begin{center}
\includegraphics[width=8cm]{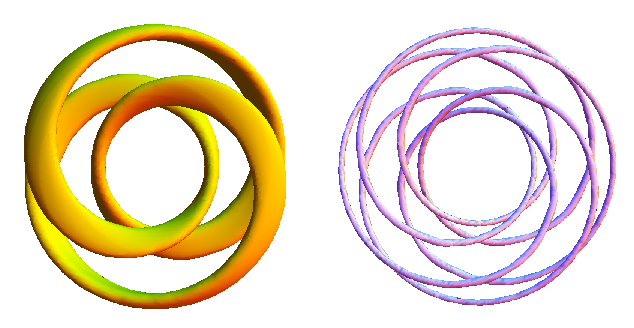}\\
\includegraphics[width=8cm]{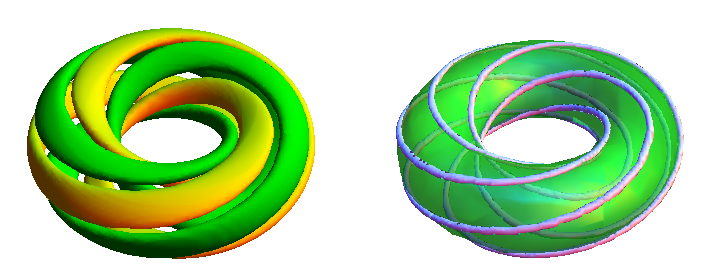}
\caption{The sixthly linked trefoil knot with winding
  numbers $(2,3)$. See the caption of Fig.~\ref{fig:32}.}  
\label{fig:23}
\end{center}
\end{figure}

The next example is the linked trefoil which is the simplest knotted
structure among the torus knots. 
Namely, they are characterized by winding numbers $(3,2)$ and $(2,3)$,
see Figs.~\ref{fig:32} and \ref{fig:23}.
The linking number is the product of $p$ and $q$. Hence, both these
trefoils are linked 6 times.
In \cite{Faddeev:1996zj}, it was conjectured that all torus knots
can be constructed as Hopfions in the Faddeev-Skyrme model, but to the 
best of our knowledge our solutions are the first ones realizing both
the $(3,2)$ and the $(2,3)$ configuration. 
For these torus knots and the remaining higher-winding ones,
%(and those given in Appendix), 
we have 
decreased the domain string width $d\simeq 1$, instead of $\sqrt{3}$
as used in the lower-winding cases, see Eq.~\eqref{eq:size}.

\begin{figure}[ht]
\begin{center}
\includegraphics[width=8cm]{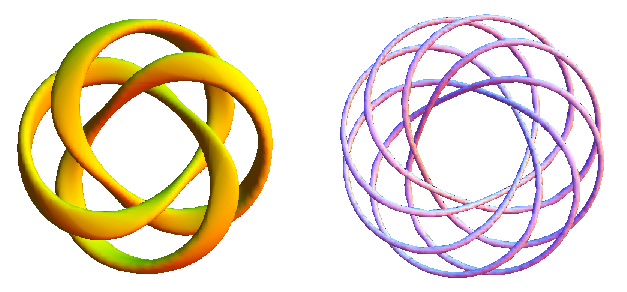}\\
\includegraphics[width=8cm]{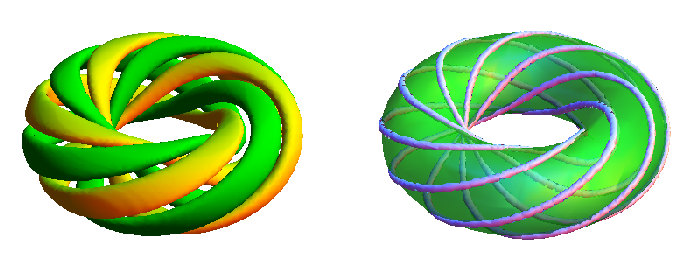}
\caption{The linked knot with winding numbers $(4,3)$.}
\label{fig:43}
\end{center}
\end{figure}

The last example is the linked knot with winding numbers $(4,3)$ and
linking number 12, see
Fig.~\ref{fig:43}. 

\section{Stability} 
There are two types of stability issues. The first is regarding the
impact of the presence of the domain strings on the host solitons;
that is, the fields have a mutual coupling that potentially could
destabilize the host soliton. This issue has been demonstrated in detail
above to depend on $\beta^2$ and be parametrically unimportant for
vanishing $\beta$. To this end, let us point out that the host soliton
is stable for $\beta=0$. Therefore, size- or shape-changing modes of
the host soliton are massive in the case of $\beta=0$. 
Let the shape-mode be gapped with mass $m^2>0$.
Once $\beta>0$ is turned on, $m$ changes accordingly and could
potentially become tachyonic. However, by continuity, an infinitesimal
change in $\beta>0$ from $\beta=0$ is not expected to be able to
drive $m^2$ negative from a finite positive value. Thus for
sufficiently small $\beta$ this type of instability is prevented. 
The other type of instability is due to the domain strings coming in
pairs of string and anti-string. We will discuss this issue below.

Since ${\cal L}_2$ is effectively the $\mathbb{Z}_2$ Wess-Zumino model
on the surface of a torus, the domain strings are characterized by
two winding numbers, $(p,q)$, and also by the $\mathbb{Z}_2$ charge. 
Each string has either $+$ or $-$ $\mathbb{Z}_2$ charge.
All configurations are topologically protected in the sense of the
topologically non-trivial winding numbers $(p,q)$. 
However, since the torus surface is periodic, a single domain string
is always accompanied by its anti-domain string. Hence, the total
$\mathbb{Z}_2$ charge is trivial. 
Therefore, all the domain-string configurations which we have obtained
in this letter are non-topological solitons.

Although their stability is granted by topology, we expect
meta-stability and thus a sufficiently long life time of the
configurations. This is due to the interaction between the
$\mathbb{Z}_2$ kinks being exponentially suppressed as $V \sim
\exp(-mL)$ where $m$ is a mass parameter 
which is of the order of the inverse kink width $d$ 
and $L$ denotes the separation distance of the
$\mathbb{Z}_2$ kink and anti-kink. 
Indeed, the asymptotic potential can be numerically calculated with
the superposition of the well-known kink and anti-kink solutions, and 
one finds that the potential has a large plateau in the asymptotic
region. 
Only when the two kinks are close enough together, they feel a finite
(non-negligible) attractive force. 
Since the two domain strings in all the constructed configurations are
nearly maximally separated, the attractive interactions among them are
small enough to render the configurations sufficiently stable. 
An exception, however, is the unknot of type $(0,1)$ which clearly is
unstable against small perturbations.

When we increase either winding number,
$p$ or $q$, a fixed size torus leaves less and less space for the
individual domain string in order not to be too close to its
anti-domain string. That is, when that happens, 
they will simply annihilate and leave behind a single vacuum. Hence,
a crude estimate of the maximum allowed winding numbers of the strings
are 
\begin{equation}
d \ll
L \sim \frac{\pi R_{\rm P} R_{\rm T}}{\sqrt{q^2R_{\rm T}^2 + p^2 R_{\rm P}^2}},
\label{eq:size}
\end{equation}
where $R_{\rm P}$ is the poloidal radius and $R_{\rm T}$ is the inner
toroidal radius. 
Here, we chose $d=v/\beta=\sqrt{3}$ (or $d\simeq 1$ in
Figs.~\ref{fig:32}--\ref{fig:43}) and $R_{\rm P,T} = {\cal O}(10)$. 

A further point in favor of the argument of stability is the 
numerical method used being a relaxation method. The relaxation method
only stops and gives a configuration when a locally stable, or
alternatively long-lived metastable configuration has been obtained. 

There are three options for improving the stability.
i) Charging the domain strings in such away that they repel
or attract each other by some kind of confining force. 
ii) A stable bound state of a string and anti-string may exist.
iii) Considering a periodic model as ${\cal L}_2$.
All of these may be realized by changing the model ${\cal L}_2$. The
third option might be the best choice. As an example, we can choose a 
modified sine-Gordon model for ${\cal L}_2$ with a periodic field \`a
la axion field. 
Because the sine-Gordon model is periodic by definition, we do not
need the anti-domain string.
Thus, the single domain string can exist by itself in such a
model. We will report on this possibility elsewhere.

\section{Conclusions} 
In this Letter, we obtained numerical solutions of new knotted domain
strings on the surface of a ring soliton in $3+1$ dimensions. We found
several torus knots with winding numbers $(1,0)$, $(2,1)$, $(3,2)$ and
$(4,3)$. With these results, we expect that all torus knots, i.e.~with
any co-prime integers $(p,q)$ can be constructed as domain strings 
in dynamical systems \footnote{See Eq.~\eqref{eq:size} for an
  approximate limit on the winding numbers of the strings as function
  of the radii of the torus.}. 
It is known that lots of complicated three-dimensional shapes can be
formed as solitons, for instance, the Buckyball was
found as a higher-charged Skyrmion. 
Even on such a complicated two-dimensional surface as host soliton, we
may construct domain strings. 
We would like to emphasize that this is indeed a doable task since the  
method of \cite{Sutcliffe:2003et} is really simple and changing the
topology of the host soliton does not lead to any difficulties.
We hope that the knotted domain strings found here will open new
research directions in many areas of physics and mathematics.  

\bigskip
{\noindent\bf Note added:} Ref.~\cite{Kobayashi:2013xoa} appeared
recently on the arXiv and contains similar torus-knots, however in a
different model.

\section*{Acknowledgments} The work of M.~E.~is supported by 
Grant-in-Aid for Scientific Research from the Ministry of Education,
Culture, Sports, Science and Technology, Japan (No. 23740226) 
and Japan Society for the Promotion of Science (JSPS) and Academy of
Sciences of the Czech Republic (ASCR)  
under the Japan - Czech Republic Research Cooperative Program. 
The work of S.~B.~G.~is partially supported by the American-Israeli
Bi-National Science Foundation and the Israel Science Foundation
Center of Excellence.

%%%%%%%%%%%%%%%%%%%%%%%%%%%%%%%%%%%%%%%%%%%%%%%%%

\end{document}